# Spatially Resolved Photo-Excited Charge Carrier Dynamics in Phase-Engineered Monolayer MoS$_2$

Hisato Yamaguchi[1#], Jean-Christophe Blancon[2#], Rajesh Kappera[3], Sidong Lei[4], Sina Najmaei[4], Benjamin D. Mangum[5], Gautam Gupta[1], Pulickel M. Ajayan[4], Jun Lou[4], Manish Chhowalla[3], Jared J. Crochet[2], Aditya D. Mohite[1*]

[1]MPA-11 Materials Synthesis and Integrated Devices, Materials Physics and Applications Division, Los Alamos National Laboratory, Los Alamos, NM 87545
[2]C-PCS Physical Chemistry and Applied Spectroscopy, Chemistry Division, Los Alamos National Laboratory, Los Alamos, NM 87545
[3]Materials Science and Engineering, Rutgers University, Piscataway, NJ 08854
[4]Materials Science and NanoEngineering, Rice University, Houston, TX 77005
[5]Pacific Light Technologies, Portland, OR 97201

#These authors contributed equally to this work.
*Corresponding author: **amohite@lanl.gov**

**ABSTRACT**

A fundamental understanding of the intrinsic optoelectronic properties of atomically thin transition metal dichalcogenides (TMDs) is crucial for its integration into high performance semiconductor devices. Here, we investigate the transport properties of chemical vapor deposition (CVD) grown monolayer molybdenum disulfide ($MoS_2$) under photo-excitation using correlated scanning photocurrent microscopy and photoluminescence imaging. We examined the effect of local phase transformation underneath the metal electrodes on the generation of photocurrent across the channel length with diffraction-limited spatial resolution. While maximum photocurrent generation occurs at the Schottky contacts of semiconducting (2H-phase) $MoS_2$, after the metallic phase transformation (1T-phase), the photocurrent peak is observed towards the center of the device channel, suggesting a strong reduction of native Schottky barriers. Analysis using the bias and position dependence of the photocurrent indicates that the Schottky barrier heights are few meV for 1T- and ~200 meV for 2H-contacted devices. We also demonstrate that a reduction of native Schottky barriers in a 1T device enhances the photo responsivity by more than one order of magnitude, a crucial parameter in achieving high performance optoelectronic devices. The obtained results pave a pathway for the fundamental understanding of intrinsic optoelectronic properties of atomically thin TMDs where Ohmic contacts are necessary for achieving high efficiency devices with low power consumption.

Graphene has attracted much attention in recent years due to its true two-dimensional (2D) geometry and one of the highest charge carrier mobility available for any material.[1] The bottleneck of its use in electronic and optoelectronic applications, however, lies in lack of an intrinsic band gap.[2] Graphene-based field effect transistors (FETs) do not turn-off completely, and the lifetimes of photo-excited carriers are too short for practical devices due to its semi-metal nature.[3] An emerging class of 2D nanomaterials, which has the complementary properties to that of graphene is transition metal dichalcogenides (TMDs). They have intrinsic band gaps in the visible range (1-2 eV) and relatively high charge carrier mobilities (200-300 $cm^2/Vs$) that are expected to overcome the limitations of graphene-based optical, electronic and optoelectronic 2D devices.[4-12] One of the key challenges in this field of research is to achieve Ohmic contacts to TMD-based semiconductor devices in order to access intrinsic material properties. Schottky contacts provide an extrinsic resistance to current flow and degrade the device performance.[13-15] One accepted process to obtain low resistance contacts to 2D TMDs such as $MoS_2$ at present is to use gold and anneal for several days under inert atmosphere.[6, 16, 17] However, its time constraints and narrow selection of electrode metals limits the breadth of potential applications for TMD-based devices.

A recent approach in achieving low resistance contacts is to partially convert $MoS_2$ semiconducting (2H) phase into metallic (1T) phase[18-21] and deposit metal electrodes selectively on the metallic 1T-phase region. We recently demonstrated the application of this phase-engineering approach to obtain high performance TMD-based FETs *via* formation of low resistance contacts that is independent of the metal used as an electrode.[22, 23] Here, we explore the effect of phase-engineered (phase-transformed) contacts on the operation of $MoS_2$ optoelectronic devices by means of scanning photocurrent microscopy (SPCM)[24-38] correlated with photoluminescence (PL) imaging. More specifically, by comparing the photocurrent profiles along the device channels of the 1T- and 2H-phase contacted devices with/without external bias, we provide a quantitative description of the reduction/elimination of Schottky barriers at the contacts, supported by proposed band diagrams to qualitatively explain the obtained SPCM results. We also analyzed the photo responsivity of devices, a figure of merit that is critical in designing minority carrier based optoelectronic devices such as solar cells and photodetectors.

Our study provides insights into efficient and optimum design of high performance TMD-based devices *via* simple and reliable procedures for forming Ohmic-like contacts to access intrinsic material properties.

**RESULTS and DISCUSSION**

*PL maps and I-V characteristics*

The photocurrent response of monolayer $MoS_2$ devices was investigated using the experimental setup illustrated in Fig. 1(a). Briefly, a chopped 440 nm laser excitation beam is focused on the sample with a spatial resolution of about 350 nm where a galvanometer mirror positioning system allows for surface rastering of the sample to obtain photocurrent maps in SPCM. The photo-response of the samples was measured by synchronous detection (at the light excitation frequency *f*) of the current at a contact of the device (drain). Devices are biased by applying a voltage potential difference $V_{SD}$ to the other contact electrode (source). Unless otherwise mentioned, the photocurrent response of 2H- and 1T-contacted $MoS_2$ devices were measured in air under ambient conditions, and with laser excitation powers below 1 μW (< 1 kW/cm$^2$). Both PL and reflection imaging capabilities were used to locate the device positions including the contact electrodes, and were correlated with SPCM results (Supporting Information Fig. S1) (see Method section for details). Fig. 1(b) depicts dark current-voltage (I-V) characteristics of 2H and 1T devices, consistent with our previous investigations. [22, 23] Specifically, while typical rectifying behavior due to a formation of Schottky barriers at the contacts was observed for 2H devices (blue line), 1T devices (red line) demonstrated a linear behavior at low applied $V_{SD}$ with enhanced current levels, indicating a strong reduction in the Schottky barrier height (or the formation of Ohmic-like contact). PL images of 2H and 1T devices (Fig. 1(c) and (d), respectively) show high intensity regions in partial triangular shapes corresponding to monolayer CVD grown $MoS_2$ sheets, as well as low intensity 'cutting' regions indicating the positions of metal electrodes. Dashed green lines highlight the position of the contacts between the $MoS_2$ channels and electrodes, matching well the optical microscopy images depicted in Fig. S1(a). PL and optical microscopy surface images show no drastic difference between 2H and 1T devices because phase converted regions of the latters are fully covered with metal electrodes by design. The devices used in SPCM measurements had channel lengths of about 5 μm (4.88 μm for the 2H device shown in Fig. 1(c) and 5.15 μm for the 1T device in Fig. 1(d)), since shorter channel length lead to complexity in data analysis due to merging of photocurrent peaks at the contacts (Supporting Information Fig. S2).

*Comparison of biased SPCM maps and line profiles*

SPCM on the 2H-contacted device at zero bias ($V_{SD} = 0$ V) presents strong photocurrent ($I_{ph}$) with opposite polarity located near the two contacts at both ends of the device channel (Supporting Information Fig. S3(a)). The result is consistent with the previous report, [39] which the photocurrent primarily originates from the Schottky barriers at the $MoS_2$/metal contacts. These SPCM features remain largely unchanged under applied bias of $V_{SD} = -0.15$ V, with only difference in the degree of $I_{ph}$ at each contact and their ratios (Fig. 1(e)). On the contrary, the 1T device at zero bias shows broad and moderate $I_{ph}$ intensity region near its channel center, in addition to relatively high Schottky photocurrent still observed at the contacts (Supporting Information Fig. S3(b)). At $V_{SD} = 0.1$ V, in stark contrast to a 2H case, we observe almost complete extinction of the photocurrent peaks at the contacts, and the appearance of a broad and high $I_{ph}$ intensity region near the channel center that spans over the entire device area (Fig. 1(f)).

The details of photocurrent evolution *versus* $V_{SD}$ for each case of a 2H and 1T device can be investigated by analyzing the line profiles of the photocurrent $I_{ph}$ across the device, that are taken with increased number of position steps compared to the SPCM maps (between the two electrodes, lines along which the profiles were taken are indicated by gray dotted lines in Fig. 1(e) and (f)). For a 2H device (Fig. 2(a)), the high $I_{ph}$ intensity regions near the $MoS_2$/metal contacts (indicated by the red triangle and blue circle) remained at same positions over the entire tested bias range ($V_{SD}$ from -0.5 to 0.5 V) while their degree increased as $V_{SD}$ increased (yellow regions indicates the position of the electrodes). For a 1T device, on the other hand, high $I_{ph}$ intensity region emerges near the center of the device channel with applied bias, and its amplitude keep increasing as the voltage increases (Fig. 2(b) top). The difference between a 2H device case is clear by comparison with the data plotted in a same $V_{SD}$ range (Fig. 2(b) bottom). Moreover, only few mV of applied $V_{SD}$ was enough to induce the high $I_{ph}$ intensity region in the channel center of a 1T device (Fig. 2(c)). The same observations are drawn from the SPCM maps depicted in Supporting Information (Fig. S4 and S5, corresponding to 2H and 1T devices, respectively). Fig. 2(d) is a comparison of photocurrent $I_{ph}$ amplitude between 1T- and 2H-contacted devices probed at the center of the channels (amplitude of a 2H device is multiplied by 100 for a demonstration purpose). The photocurrent level under bias is much stronger in 1T devices compared to a 2H case, which is consistent with the obtained dark I-V characteristics (Fig. 1(b)). These results clearly demonstrate that a use of devices with 1T contacts over 2H devices allow wider tunability of the active area with enhanced photocurrent response in monolayer $MoS_2$ based optoelectronic devices. Error bar for photocurrent $I_{ph}$ was higher for a 1T-contacted device compared to a 2H case, however, it was still well below 10%. The larger error bar for a 1T device is possibly due to much reduced Schottky barrier heights at its contacts, which enables a device to be more sensitive to small local potential fluctuations compared to a 2H device.

*Schottky barrier height analysis using SPCM line profiles*

Analysis of the obtained $I_{ph}$ line profiles can provide insights into quantitative values of Schottky barrier height (SBH) at the contacts. In our analysis, $V_{SD}$ dependence of the photocurrent profiles along the channels of 2H and 1T devices were plotted for the following three signature points: positions near each $MoS_2$/electrode contacts ($I_{ph,S}$ and $I_{ph,D}$ at the source and drain electrodes, respectively) and at the center region ($I_{ph,C}$) - as indicated by red triangles (source), blue circles (drain), and green squares (center) in Fig. 2(a), (c) respectively. The results are plotted in Fig. 3(a) for a 2H device (corresponding to Fig. 2(a) $I_{ph}$ profiles), and in Fig. 3(b) for a 1T device (extracted from the $I_{ph}$ response including that of Fig. 2(c)).

In a 2H device case (Fig. 3(a)), $I_{ph,S}$ and $I_{ph,D}$ are nearly symmetric with respect to the origin ($I_{ph}$ = 0 nA and $V_{SD}$ = 0 V), where $I_{ph,S}$ ($I_{ph,D}$) shows almost linear increase with positive (negative) $V_{SD}$. More precisely, extrapolation of $I_{ph,S}$ ($I_{ph,D}$) between 0.5 and -0.15 V (-0.5 and 0.05 V) as indicated in a red (blue) dashed line intersects the Y-axis ($I_{ph}$ = 0) around -0.21 V (0.21 V), which provides a quantitative estimate of the SBH at the source (drain) contact because the observed change in polarity of $I_{ph}$ at the contacts can be explained by the $V_{SD}$ exceeding their heights of the built-in potential (Schottky barrier). [25, 29, 30] A further discussion is presented later.

The origins of quasi-polarity change (not a complete change) observed for $I_{ph,S}$ and $I_{ph,D}$ in our case could be due to external effects such as environmental doping or Fermi level pinning. [13, 14] SBH of ~210 meV, extracted from our SPCM results is consistent with a true (effective) SBH extracted from variable low temperature electrical characterizations (flat band measurements) [13] that we performed independently using the same electrode metal on $MoS_2$. The photocurrent at the channel center $I_{ph,C}$ increases monotonically over the entire range of applied bias at a smaller amplitude ($I_{ph,S}$ and $I_{ph,D}$ are measured in the 0.1 nA range whereas $I_{ph,C}$ operates at a level of 0.01 nA due to absence of a Schottky barrier) and crosses the Y-axis around -0.16 V (as indicated by a green line), which is in the range of the intersects observed for $I_{ph,S}$ and $I_{ph,D}$ (gray region on Fig. 3(a)). The agreements between $I_{ph,C}$ intersect and that of the $I_{ph,S}$ and $I_{ph,D}$ can be understood by considering that $I_{ph,C}$ intersect providing 'average' SBH of the whole device. The slightly different behaviors between the Schottky barriers at each contact is possibly due to a combination of minor difference in the sharpness of the contact $MoS_2$/electrode interface created during the electrodes deposition, [13] and intrinsic/extrinsic inhomogeneity of work function over the $MoS_2$ surfaces (created during the CVD growth/external environment effects such as water molecule absorptions, respectively)). [40, 41]

Two major differences observed in a 1T device case (Fig. 3(b)) from that of a 2H case were an order of magnitude larger total $I_{ph}$ for the same bias range, and Y-axis intersect of $I_{ph}$ at almost zero bias. A more details of the higher amplitude of $I_{ph}$ is discussed later but it is consistent with dark I-V characteristics (Fig. 1(b)) as mentioned earlier for the case of Fig. 2(d). For a precise determination of Y-axis intersect of $I_{ph}$, $I_{ph}$ at marked three positions (two contacts and center of the device channel) were plotted in much smaller range of bias in Fig. 3(b) (between -10 and +10 mV) compared to a 2H case (Fig. 3(a)). $I_{ph,D}$ and $I_{ph,S}$ cross the Y-axis at 5.8 mV and -6 mV, as indicated in a red and blue dashed line respectively, providing quantitative values for the reduced SBHs after its 1T-phase transformation (obtained SBH value range is colored in gray). $I_{ph,C}$ crossed Y-axis at zero (as indicated by a green line) within the resolution of the measurements (< 1 mV), which can be interpret as a 'average' of source and drain Schottky barriers similar to the case of a 2H device. Note that the obtained value of $I_{ph,C}$ being very close to the average between the two SBHs (-0.1 mV) could be an indication that this particular device had well-balanced source and drain Schottky barriers (having identical heights). The validity of the obtained results was also confirmed on other 1T devices, including the one which we performed photocurrent microscopy at a single position to achieve site specific SBH values (Supporting Information Fig. S5, S6). In this case, the SBH was estimated by illuminating the device at a fixed position while sweeping the bias voltage and recording $I_{ph}$ (Supporting Information Fig. S6(b)). Locating the laser spot at an appropriate position near the contact allows a determination of SBH with improved accuracy (spatial and energy) with shorter measurement time compared to the line or map scan counterparts. Analysis of the data yields ~0.05 and ~0.46 meV for the drain and source SBH, respectively, reaffirming the strong reduction of SBH after a phase transformation to 1T. The photocurrent at the channel center crosses the Y-axis at ~0.37 mV, indicating that the source Schottky barrier dominates the response of this particular device at small bias range due to its larger amplitude compared to the drain. All SBH values we obtained for 1T devices are up to two/three orders of magnitudes lower than a 2H device case, quantitatively demonstrating the strong reduction of native SBH and hence the formation of nearly Ohmic-like contacts for monolayer $MoS_2$ using the phase engineering approach (*i.e.* by converting of the $MoS_2$ 2H-phase into 1T-phase underneath the contact electrodes). An effect of illumination intensity on a determination of SBH should be negligible in this study, unlike a case of open circuit voltage $V_{OC}$ for organic photovoltaics. This is due to four orders of magnitudes higher carrier mobility in our $MoS_2$ (~10 $cm^2$/Vs) [22, 23] compared to that of an organic photovoltaic system (~$10^{-3}$ $cm^2$/Vs), [42] which reduces the degree of illumination intensity dependence on SBH by at least several orders of magnitudes (see Supporting Information S7 for details). [43, 44]

*Insights into potential profile across devices*

Complementary qualitative analysis of the Schottky barriers was obtained by plotting the integrated photocurrent along the devices channels (Fig. 4(a), (b) derived from Fig. 2(a), including that of (c), respectively - see also Supporting information Fig. S7), that provides insight into potential profile of 2H and 1T devices. Although this method does not provide direct access to the intrinsic potential profile across the device as suggested in ref. 45, it offers useful qualitative information on the relative evolution of the potential profile depending on the applied bias. In a 2H device case (Fig. 4(a) and Supporting Information Fig. S7(a)), strong variations of the potential profiles take place at the source (drain) contact position for positive (negative) bias, indicating strong build-in local electric field drives the separation of charges and thus enhancing the photocurrent. The results are in agreement with formation of SBH at the contact as observed in Fig. 2(a), 3(a). On the other hand, a 1T device (Fig. 3(b) and Supporting Information Fig. S7(b)) shows a much smoother potential modulation across the device channel. These observations validate the absence of significant build-in field along the channel and confirm the strong attenuation of the Schottky barriers in 1T devices, as quantitatively discussed above.

*Photocurrent generation mechanism*

Fig. 5(a)-(d) illustrates a possible mechanism, which explains the observed photocurrent generation in 2H- and 1T-contacted devices of monolayer $MoS_2$. Specifically, the conduction and valance band potential profiles across the device channel are drawn along with our interpretation of the charge diffusions, in correlation with the schematics of the photocurrent line profiles achieved in the SPCM. For our explanations, we classified the generated photocurrent into the following two dominant types; a Schottky barriers driven photocurrent located at $MoS_2$/electrode contacts (noted $I_{SC}$), and a combination of all the other types of photocurrent $I_{PC}$ [38, 45-48] (most likely dominated by photoconductive photocurrent as indicated in ref. 48). We believe that a thermoelectric effect has only a minor contribution to our observed overall photocurrent as confirmed by recent studies (details are provided in Supporting Information S9). [39, 48] At zero bias ($V_{SD}$ = 0 V) and under illumination, 2H and 1T devices yield dominant photocurrent $I_{SC}$ at both of source and drain contacts due to presence of Schottky barriers. These Schottky barriers generate a local built-in electric field that separate the photo-excited carriers at the contacts (residue SBH for 1T case), and drives the hole (electron) carriers towards (away from) the closest electrode (carrier flows depicted by red arrows in Fig. 5(a), (c)). The depletion width of Schottky barriers is much smaller than the spatial resolution of our SPCM setup (~50 nm [45] *versus* few hundreds of nm), thus results in relatively narrow high intensity peak at the contacts. In this case, an amplitude of the photocurrent generated at the contacts is determined by the amplitudes of SBH thus resulting in higher $I_{SC}$ for a 2H case compared to a 1T case (Fig. 5(a) and (c), respectively). This is consistent with our SPCM results without any bias ($V_{SD}$ = 0 V) that yielded an order of magnitude higher $I_{ph}$ for a 2H device compared to a 1T device due to a difference in their SBH (~200 meV and < 10 meV for 2H and 1T, respectively). On the other hand, the photo-excited carriers located further away from the contacts/towards center region of a 2H device do not contribute to $I_{ph}$ due to lack of a driving potential. In a 1T device, however, the moderate number of photo-generated carriers can reach the electrodes due to absence/reduction of potential barriers at the contacts, and contribute to photocurrent $I_{ph}$ observed at the device channel center. This interpretation suggests that the broad and moderate $I_{ph}$ feature in the center of the device can be used as an important indication of dramatic reduction of Schottky barrier.

Under a bias, in a positive case for example, above mentioned phenomena appears to be pronounced and emphasizes the differences in SPCM response between 2H and 1T devices (the same analysis is possible for negative bias case by exchange of source and drain contacts). In a 2H device (Fig. 5(b)), applied external field affects the SBH at the both contacts. On the source, it enables carriers that are generated at positions further away from the source to reach electrodes (source for holes and drain for electrons) and contribute to $I_{SC}$ as indicated by longer red arrows. On the drain, however, it decreases $I_{SC}$ at the drain contact in relative to the source due to decreased SBH (indicated by shorter arrows). Positions of high $I_{SC}$ regions remain unchanged. It should be noted that this analysis is valid in a bias range that is away from the saturation regime of the device, and 0.5 V used in our case is well below the critical voltage. [23] In 1T devices, the elimination effect of Schottky barriers becomes even more apparent upon applying an extremely small bias of few mV (Fig. 5(d)). The tilt/bending of the band potential rapidly enable photo-excited carriers to overcome the SBH at both contacts and render $I_{SC}$ negligible (the peaks located at the contacts disappear). This allows for uninhibited flow of photo-excited carriers even under very small external electric-fields unlike in the case of 2H-contacted devices. As a result, $I_{PC}$ maximized near the center of the channel emerges and spans over the entire device. This would not be possible with 2H-contacted device because a formation of Schottky barriers at the electrodes hinders electrons from reaching the electrodes, resulting in no observable total photocurrent $I_{ph}$ (Fig. 5 (b)). The gate dependence of SPCM results is expected to provide further insights into the evolution of Schottky barrier heights and detailed photo-generation mechanism in $MoS_2$ based devices, and should be investigated in future works.

**Device performance**

Integrated photocurrent response along device channels (Fig. 6(a)) and the local photo responsivity (Fig. 6(b)) demonstrate the enhanced performances of 1T-contacted devices in comparison to their 2H counterparts. We observe same trends in the integrated photocurrent and dark current variations as function of the bias (Fig. 2(a)), as well as comparable ratio of current amplitudes between the 2H and 1T devices (about one order of magnitude difference). This is consistent with the trend of 133% carrier mobility increase observed for 1T device FETs in our previous reports. [22, 23] Moreover, a 1T device presents photo responsivity R (ratio of the local $I_{ph}$ to the excitation power) about 30 times larger (R=5.5 mA/W) than a 2H device case (R=0.2 mA/W) at $V_{SD}$ = -0.5 V. A 2H device responsivity is almost invariant over the whole $V_{SD}$ bias range of -0.5 to 0.5 V, in agreement with the fact that $I_{SC}$ (Schottky barriers driven photocurrent) dominates the photo-response. On the contrary, a 1T device shows strong responsivity variations in a V-shape centered at $V_{SD} \approx 0$, where R presents a drastic increase as $V_{SD}$ overcome the SBH (photocurrent regime changes from $I_{SC}$ at low bias (-10 mV $\leq V_{SD} \leq$ 10 mV) to $I_{PC}$). The responsivity R = 6.3 mA/W at $V_{SD}$ = 0.5 V (excitation power 0.7 µW at 2.8 eV (440 nm)) is larger than observations in monolayer graphene transistors [47] and comparable to recent reports in monolayer MoS$_2$ under similar operation parameters. [48-50] We expect further enhancement with illumination at the excitonic absorption states. [5] Our results demonstrate effectiveness of using the phase engineering approach (conversion of 2H- to 1T-phase) for monolayer MoS$_2$ to eliminate the Schottky barriers at the contacts, thus improving the operating current level as well as increasing the net active area of the device throughout the entire channel to achieve high performance 2D optoelectronic devices.

**CONCLUSION**

In conclusion, we performed SPCM on monolayer MoS$_2$ individual sheet devices with phase-engineered contacts and compared the results with conventionally used 2H-contacted devices. The results revealed that in addition to the increased photocurrent level, the active area of monolayer MoS$_2$ optoelectronic devices broadened over their entire area due to Schottky barrier elimination of a 1T device. This is in stark contrast to narrow active regions near the electrodes for a 2H device. Furthermore, analysis of SPCM results indicated that the Schottky barrier heights (SBH) at the contacts of a 1T device is reduced by at least one order of magnitude compared to a 2H case, from ~200 meV down to few meV or even lower. Our proposed model suggests that this elimination of Schottky barriers at the electrodes achieved by the conversion of semiconducting 2H-phase to metallic 1T-phase is responsible for the observed broadening of the active area and increase of photocurrent level. We also demonstrated that the photo responsivity increases by more than an order of magnitude for a 1T device compared to a 2H case, a promising indication towards high performance optoelectronic devices. Our results pave a pathway for the design of high performance TMD-based 2D optoelectronic devices and fundamental understanding of their optoelectronic properties.

**METHODS**

**Fabrication procedures of 1T- and 2H-phase contacted MoS$_2$ devices**

Monolayer CVD MoS$_2$ sheets [40] were transferred [51] onto degenerately p-doped (R< 0.0015Ω-cm) patterned silicon substrates capped with 100 nm oxide layer. Statically dispensed PMMA (A4, Microchem corp.) was spin coated onto the sample at 4,000 RPM for 60 seconds and was followed by pre-baking at 180°C for 90 seconds. For fabrications of 1T-phase contact devices, two separate lithography processes were performed. In the first lithography step, electrode windows were opened on the MoS$_2$ sheet using conventional e-beam lithography. After opening the electrode windows, we exposed the samples to n-butyl lithium (1.6M, Sigma Aldrich) for 48 hours and then cleaned them with hexane. [22, 23] All butyl lithium exposure was conducted in Argon filled glove box. Samples were then cleaned with deionized water in order to remove residual lithium on MoS$_2$. This was followed by PMMA etching using HPLC grade acetone (Fisher Scientific), followed by an isopropanol rinse. PMMA was again spin coated following the same recipe as earlier. Second lithography step was performed to open the electrode windows at the same regions as earlier.

We performed e-beam evaporation to deposit titanium (5nm) and gold (50 nm) under high vacuum conditions of $10^{-7}$ Torr at a slow deposition rate of 1 Å/s. This was followed by lift off using acetone after which the samples were properly rinsed with isopropanol to eliminate acetone residue followed by blow drying with compressed nitrogen gas. For fabrications of conventional 2H-phase contact devices, metal electrodes were deposited without n-butyl lithium exposure steps.

**Scanning photocurrent microscopy (SPCM) and Photoluminescence (PL) measurements**

For SPCM measurements, 440 nm pulsed excitation light was delivered by a laser diode (PicoQuant), and focused on the sample by a 60x Olympus objective with 0.9 numerical aperture to be scanned over the samples in a form of map / line / fixed position by means of a galvanometer mirror positioning system and two lenses arranged in a 4f-configuration. [34] As illustrated in Fig. 1 (a), synchronous detection was achieved *via* SR-830 lock-in amplifier (Stanford Research Systems) using the reference frequency input from the chopper (~370 Hz). The photocurrent signal was first amplified by a SR-570 current amplifier (Stanford Research Systems) and the AC component of the output voltage signal was used as direct input to the lock-in amplifier. Both the amplitude $R$ and the phase $\varphi$ of the photocurrent were monitored and analyzed. Correlated PL and/or reflected light obtained for the same position of SPCM map was detected using an avalanche photodiode and a silicon photodiode, respectively, allowing determination of the exact location of monolayer $MoS_2$ sheet devices and the contact electrodes.

**ACKNOWLEDGEMENTS**

Authors acknowledge S.E.Yalcin of Los Alamos National Laboratory (LANL) for the technical supports and J.A.Garcia of LANL for the administrative supports at the initial stage of SPCM system setup. Authors also acknowledge J.K.Baldwin of LANL for metal electrode depositions, D.J.William of LANL for maintenance of the SEM system for e-beam lithography, and W.Nie of LANL for fruitful discussions. H.Y. acknowledges financial supports from LANL Director's Postdoctoral Fellowship, and S.K.Doorn of LANL for an initial support as a host. This work was financially supported by LANL Laboratory Directed Research and Development (LDRD) program, and was performed, in part, at the Center for Integrated Nanotechnologies, an Office of Science User Facility operated for the U.S. Department of Energy (DOE) Office of Science by Los Alamos National Laboratory (Contract DE-AC52-06NA25396) and Sandia National Laboratories (Contract DE-AC04-94AL85000).

***Supporting Information Available:*** Optical, reflection, and PL images for determination of device positions, SPCM on short channel length devices, Amplitude and phase information extracted from SPCM maps, Bias dependent SPCM maps of a 2H-contacted device, Reproducibility of bias dependent SPCM maps of a 1T-contacted device, Fixed position photocurrent microscopy of a 1T-contacted device, Effects of illumination intensity on SBH determinations, Photocurrent integrated from opposite direction and Thermoelectric effect contributions in photocurrent measured in this study. This material is available free of charge *via* the Internet at http://pubs.acs.org.

**FIGURE CAPTIONS**

**Fig. 1**: (a) Schematic of the scanning photocurrent microscopy (SPCM) experimental setup used in this study. (b) Dark current-voltage characteristics of 2H- and 1T-contacted devices (blue and red curves, respectively). (c) Photoluminescence (PL) image of a 2H device and (e) its SPCM map under -0.15 V bias. (d) PL image of a 1T device and (f) its SPCM map under 0.1 V bias. Green dashed lines correspond to the position of the contacts between $MoS_2$ channels and metal electrodes. Photocurrent profiles shown in Fig. 2 are measured along the vertical gray dashed lines. S and D boxes indicate the position of the source and drain electrodes, respectively. Scale bars = 3 μm.

**Fig. 2**: (a) Photocurrent profile along a 2H device channel for different bias $V_{SD}$ in both polarities. Yellow regions indicate the positions of the source (S) or drain (D) contacts. (b) Comparison of photocurrent profiles for 1T- (top) and 2H-contacted devices (bottom) at different $V_{SD}$ (only for positive polarity for simplicity). (c) Photocurrent profile for a 1T device at low $V_{SD}$ (< 10mV) in both polarities. (d) Comparison of photocurrent amplitude between 1T- and 2H-contacted devices at the center of device channels. Amplitude of a 2H device is multiplied by 100 for a demonstration purpose.

**Fig. 3**: (a) and (b) are photocurrent amplitude *versus* applied bias $V_{SD}$ at signature points of the profiles in Fig. 2 (a) (2H-) and Fig. 2 (c) (1T-contacted device), respectively. Gray regions encompass the $V_{SD}$ range, which corresponds to Schottky barriers heights at the source and drain.

**Fig. 4**: Integrated photocurrent along (a) 2H and (b) 1T device channels for different applied bias $V_{SD}$.

**Fig. 5**: Proposed mechanisms of photocurrent generation in (a), (b) 2H-contacted and (c), (d) 1T-contacted devices. Top panels (a), (c) and bottom ones (b), (d) correspond to situations at zero and positive bias, respectively. In each panel, the top schematic illustrates the conduction (blue) and valance (green) bands variations along the device channel - the $MoS_2$ monolayer is sandwiched between the two metal electrodes (positions of the electrodes are shown in yellow,

with S and D corresponding to the source and drain, respectively). Insets illustrated enlarged MoS$_2$/electrode contact and depict the Schottky barrier height derived from our SPCM measurements. Diffusion of electrons (-) and holes (+) are represented by red arrows, which the corresponding plot schematics for relative photocurrent *versus* the position along the channel are illustrated below. The direction and relative amplitude of currents are indicated by arrows in-between the two schemes, with I$_{SC}$ indicates the photocurrent generated by the Schottky barriers and I$_{PC}$ indicates photocurrent, which is a sum of all the other types.

Fig. 6: 2H- *vs.* 1T-contacted devices performances and their evolution with applied source-drain bias V$_{SD}$. (a) Total photocurrent integrated over the channel length of 2H (black dots, right scale) and 1T (red squares, left scale) devices. (b) Local photo responsivity extracted from the SPCM measurements as the ratio of maximum photocurrent to the excitation power.

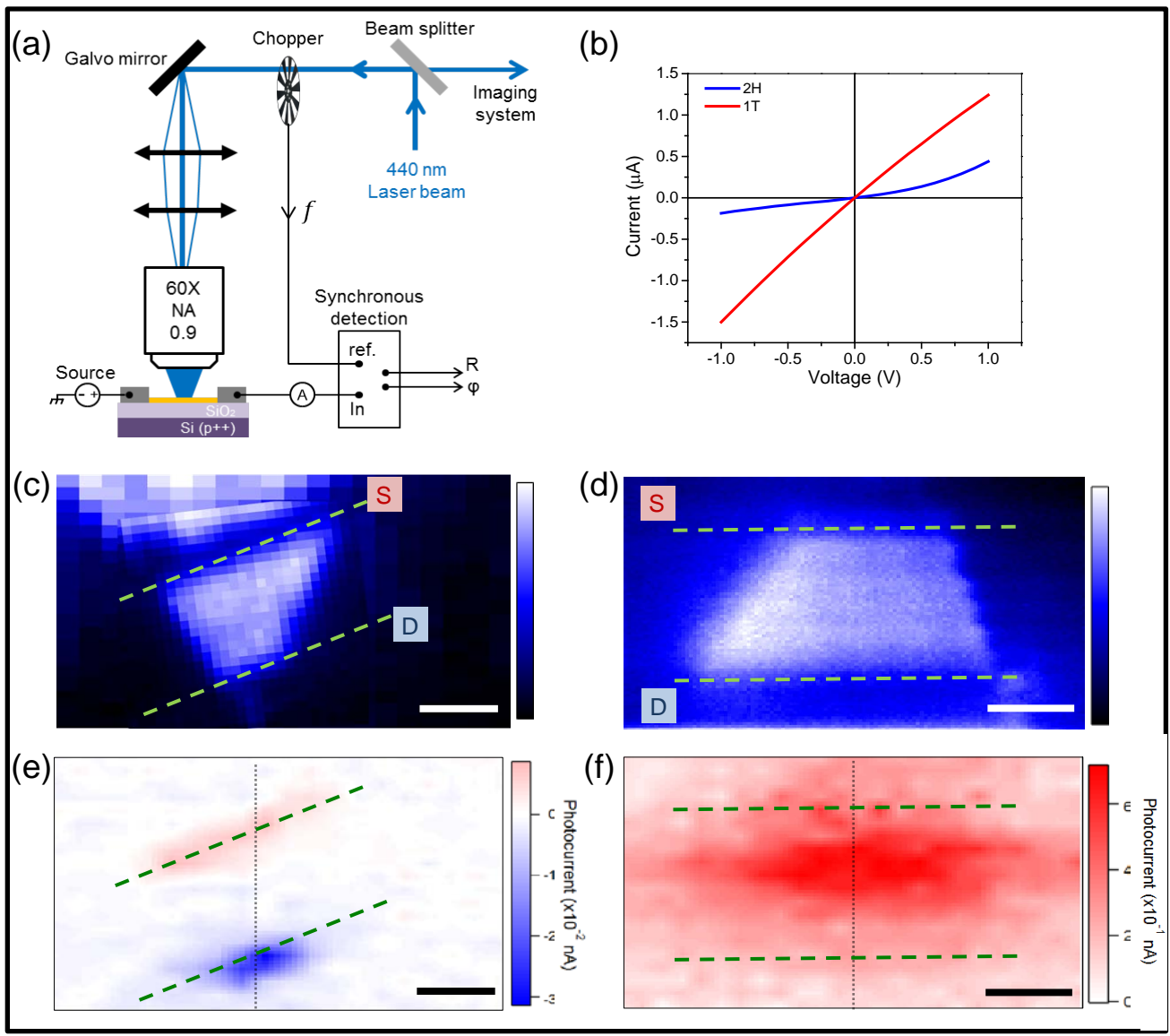

Fig. 1 H.Yamaguchi/J.-C.Blancon *et al.*

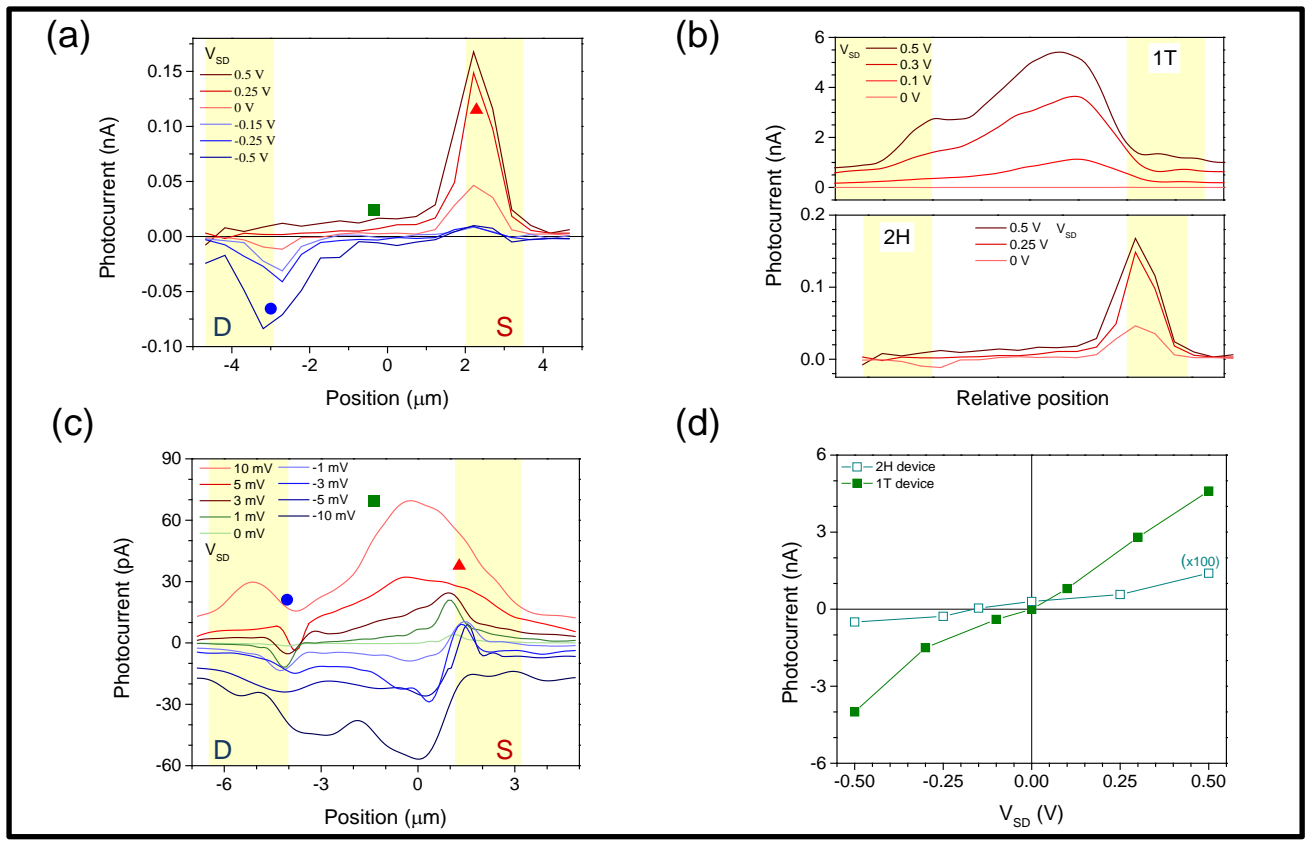

Fig. 2 H.Yamaguchi/J.-C.Blancon *et al.*

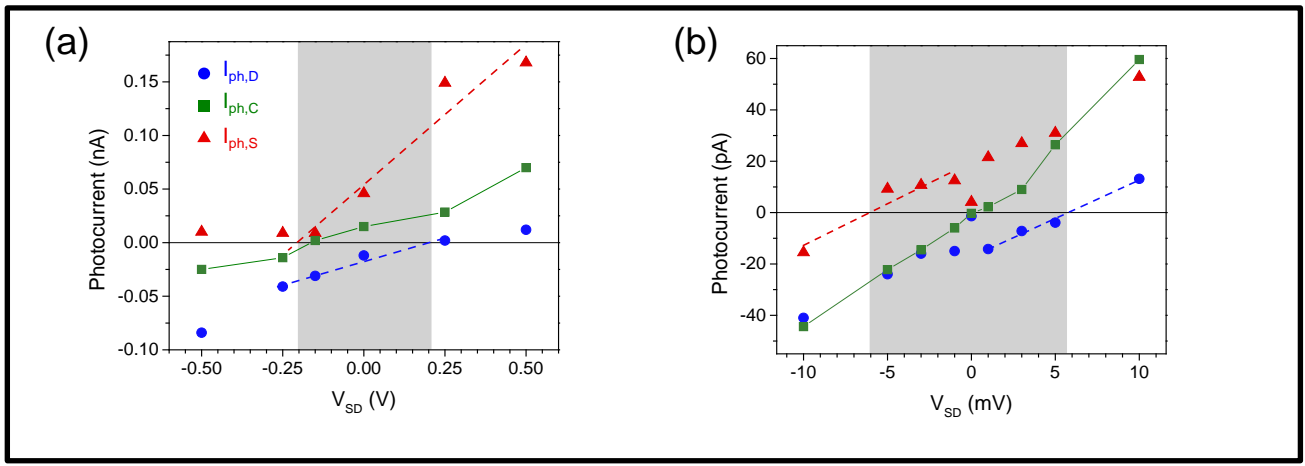

Fig. 3 H.Yamaguchi/J.-C.Blancon *et al.*

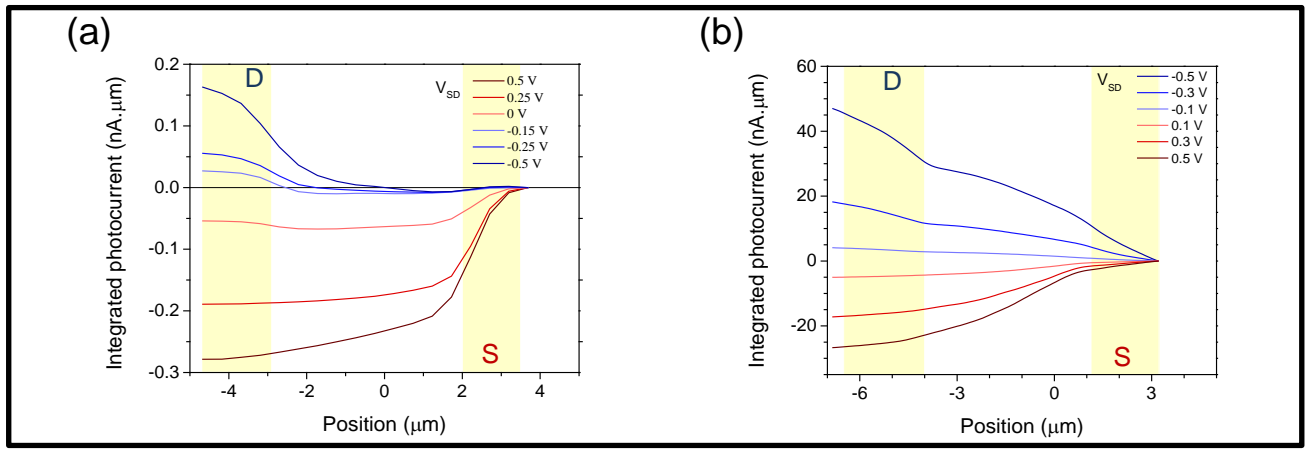

Fig. 4 H.Yamaguchi/J.-C.Blancon *et al.*

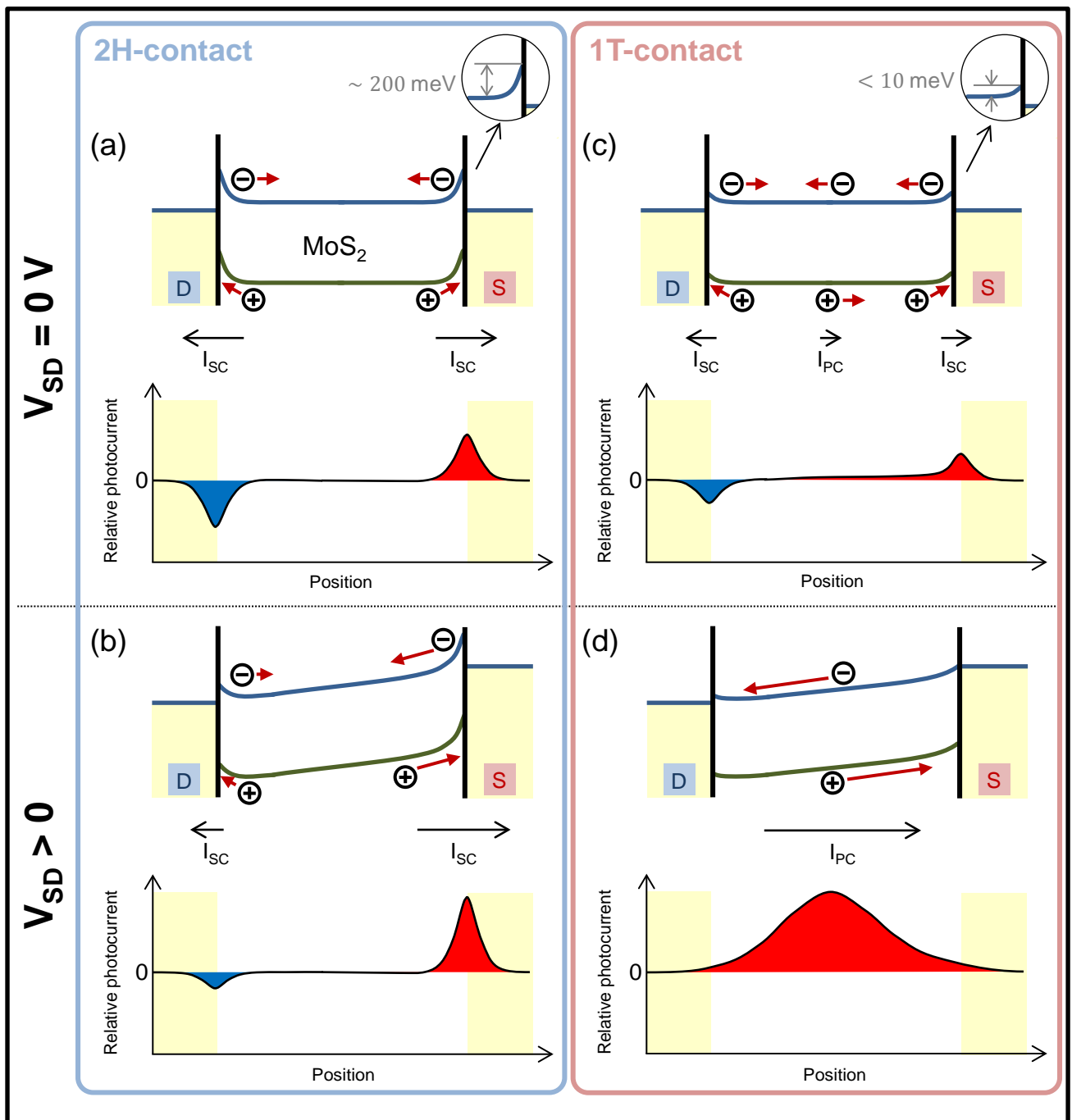

Fig. 5 H.Yamaguchi/J.-C.Blancon *et al.*

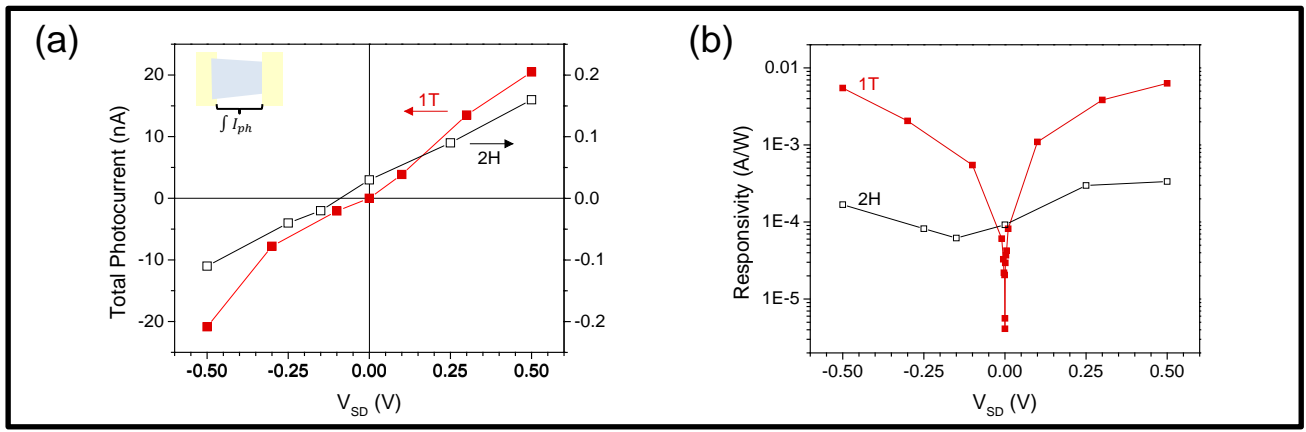

Fig. 6 H.Yamaguchi/J.-C.Blancon *et al.*

Supporting Information for

# Spatially Resolved Photo-Excited Charge Carrier Dynamics in Phase-Engineered Monolayer MoS$_2$

Hisato Yamaguchi[1#], Jean-Christophe Blancon[2#], Rajesh Kappera[3], Sidong Lei[4], Sina Najmaei[4], Benjamin D. Mangum[5], Gautam Gupta[1], Pulickel M. Ajayan[4], Jun Lou[4], Manish Chhowalla[3], Jared J. Crochet[2], Aditya D. Mohite[1*]

[1]MPA-11 Materials Synthesis and Integrated Devices, Materials Physics and Applications Division, Los Alamos National Laboratory, Los Alamos, NM 87545
[2]C-PCS Physical Chemistry and Applied Spectroscopy, Chemistry Division, Los Alamos National Laboratory, Los Alamos, NM 87545
[3]Materials Science and Engineering, Rutgers University, Piscataway, NJ 08854
[4]Materials Science and NanoEngineering, Rice University, Houston, TX 77005
[5]Pacific Light Technologies, Portland, OR 97201

#These authors contributed equally to this work.
*Corresponding author: **amohite@lanl.gov**

**Contents**
**S1 Optical, reflection, and PL images for determination of device positions**
**S2 SPCM on short channel length devices**
**S3 Amplitude and phase information extracted from SPCM maps**
**S4 Bias dependent SPCM maps of a 2H-contacted device**
**S5 Reproducibility of bias dependent SPCM maps of a 1T-contacted device**
**S6 Fixed position photocurrent microscopy of a 1T-contacted device**
**S7 Effects of illumination intensity on SBH determinations**
**S8 Photocurrent integrated from opposite direction**
**S9 Thermoelectric effect contributions in photocurrent measured in this study**

# S1 Optical, reflection, and PL images for determination of device positions

Fig. S1(a) indicate optical microscopy images of the 2H (top) and 1T (bottom) devices presented in Fig. 1(c), (d) of the main text, respectively. White dashed lines are contacts between the electrodes and monolayer $MoS_2$ devices, and "S" ("D") indicates source (drain). Fig. S1(b) is photoluminescence and (c) is reflection images of a same device, showing the partial triangular shaped CVD $MoS_2$ sheets (bright in PL and dark in reflection) as well as the metal electrodes (bright in reflection and not visible in PL). Fig. S1(d) is photoluminescence and reflection profiles along the vertical red dashed lines depicted in (b), (c). Yellow regions indicate the position of the metal electrodes. All of the scale bars are 3 μm.

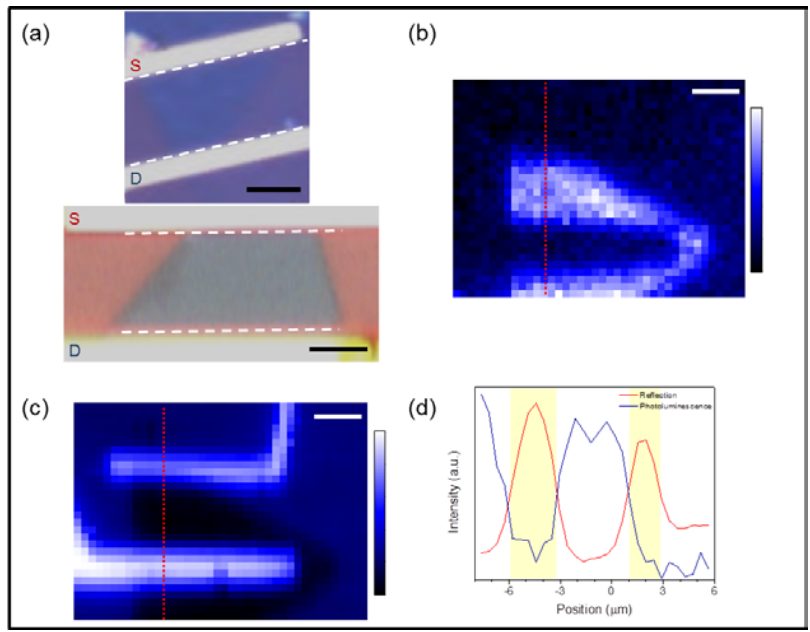

Fig. S1

# S2 SPCM on short channel length devices

Fig. S2 is SPCM measurement results for device channel lengths shorter than ~5 μm (2H devices). Top and bottom panels show different devices with channel length of 1.95 μm and 0.65 μm, respectively. For each device, (a), (c) are photoluminescence images, and (b), (d) are SPCM maps at -1 and +1 volt bias. Green dashed lines correspond to the position of the contacts between $MoS_2$ channels and metal electrodes. Scale bars for (a), (b) are 3 μm, and for (c), (d) are 1.5 μm.

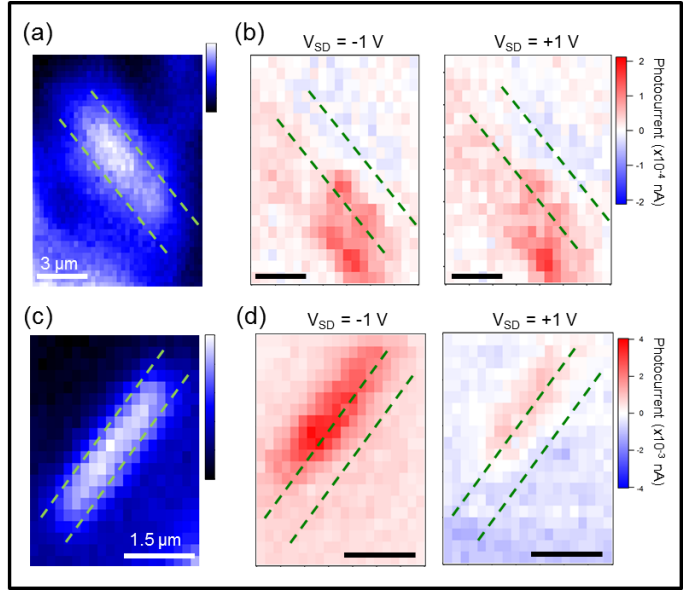

Fig. S2

**S3 Amplitude and phase information extracted from SPCM maps**

Fig. S3 is amplitude and phase information extracted from SPCM maps for a (a) 2H- and (b) 1T-contacted device at zero bias ($V_{SD} = 0$ V), respectively - for the same devices introduced in Fig. 1 of the main text. The photocurrent responses including its sign (top) are derived from both the photocurrent amplitude (middle) and phase (bottom) (R and φ signal respectively from the lockin amplifier). A ~180 deg. phase change indicates opposite photocurrent signs. Phase variations smaller than ~20 deg. are not reflected in the analysis because they usually arise from small changes of capacitances in the system. Green and yellow dashed lines correspond to the position of the contacts between $MoS_2$ channels and metal electrodes. Scale bars are 3 μm.

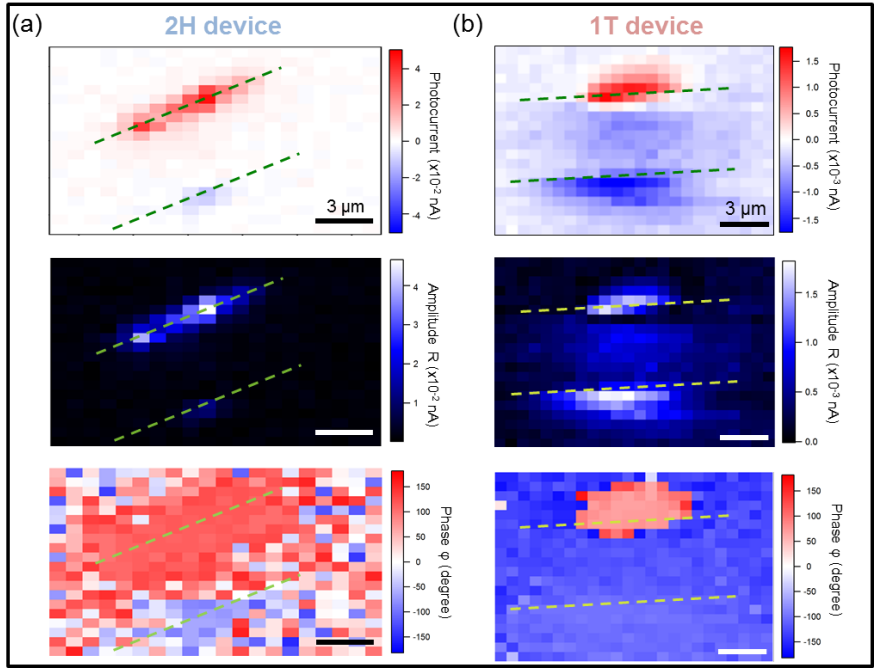

Fig. S3

## S4 Bias dependent SPCM maps of a 2H-contacted device

Bias $V_{SD}$ dependence of SPCM response of the 2H-contacted devices in Fig. 1(c) of the main text is shown in Fig. S4. Green dashed lines indicate the positions of the two MoS$_2$/metal electrode junctions. Scale bars are 3 μm.

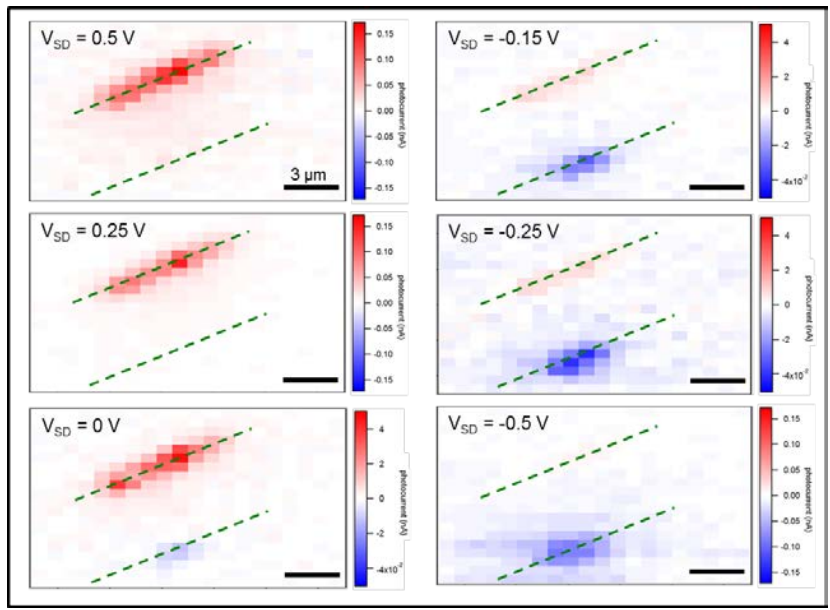

Fig. S4

## S5 Reproducibility of bias dependent SPCM maps of a 1T-contacted device

Bias $V_{SD}$ dependence of SPCM response for a different 1T-contacted device from that of in the main text is shown in Fig. S5 to demonstrate its reproducibility. The top left image is a PL map of the device, and "S" ("D") indicates source (drain). White and green dashed lines indicate the positions of the two $MoS_2$/metal electrode junctions. All of the scale bars are 3 μm.

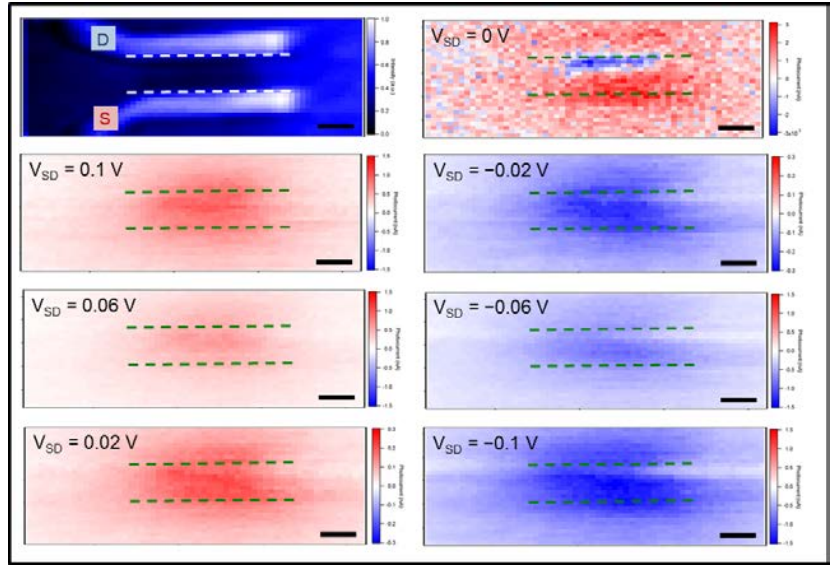

Fig. S5

## S6 Fixed position photocurrent microscopy of a 1T-contacted device

Fig. S6(a) is SPCM map of the 1T device introduced in Fig. S5 at zero bias. Green dashed lines indicate the positions of the two $MoS_2$/metal electrode junctions. (b) is bias dependence of the photocurrent probed at the three points shown in blue, green, and red circles of (a) (see main text for details). The gray region indicate range of X-axis intersects ($I_{ph}$ = 0 nA) for the three measured positions.

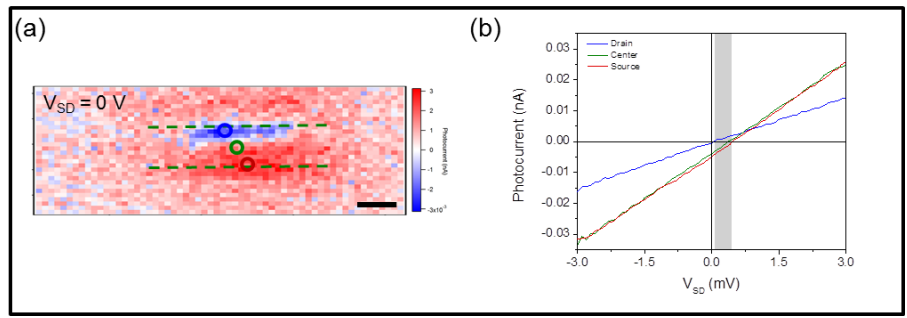

Fig. 6

## S7 Effects of illumination intensity on SBH determinations

Due to some conceptual overlaps in determination of $V_{OC}$ in a solar cell and SBH using SPCM, it might be worthwhile comparing those two cases when discussing about a possible effect of illumination intensity in determination of SBH in this study. The origin of illumination intensity dependence of $V_{OC}$ in an organic solar cell is understood by a bimolecular recombination process [1]. A bimolecular recombination is a process which free charge carriers in the materials (electrons & holes) recombine without an assist of structural defects such as trap states in a bulk or surface/interfacial states [2-3]. The process is known to be dominated by a mobility of a *lowest* carrier in the system, which a lower mobility leads to higher recombination rate hence larger illumination intensity dependence of extracted energy barrier heights ($V_{OC}$ and Schottky barrier for solar cell and our case, respectively) [4]. Specifically, a mobility of lowest carrier in an organic solar cell case (P3HT:PCBM blends for example) typically ranges in $10^{-3}$ cm$^2$/Vs [5] whereas in monolayer CVD MoS$_2$ as in our case ranges in 10 cm$^2$/Vs [6, 7]. Given a degree of $V_{OC}$ variation in an organic solar cell is ~0.2 V between 1 & 100 mW/cm$^2$ [1], an illumination intensity effect on determination of Schottky barrier heights in our case is expected to be <few eV due to several orders of magnitudes higher carrier mobility of monolayer CVD MoS$_2$ compared to P3HT:PCBM blends. Based on a discussion above, we believe that it is reasonable to assume that the effect of illumination intensity on a determination of SBH is negligible, hence our technique can serve as a reliable method for determination of energy barrier heights in a device/system under the experimental conditions we used.

## S8 Photocurrent integrated from opposite direction

Fig. S7 is photocurrent integrated along (a) 2H and (b) 1T devices channels for different applied bias from opposite direction (drain to source), complementary to Fig. 4 (a), (b) in the main text. "S" ("D") indicates source (drain), and yellow regions indicate the position of the metal electrodes.

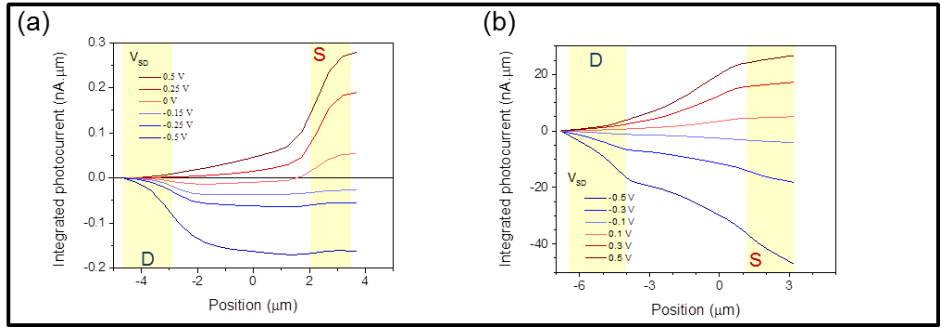

Fig. S7

**S9 Thermoelectric effect contributions in photocurrent measured in this study**

We believe that thermoelectric effect has only a minor contribution to our observed overall photocurrent for following two reasons. (1) Generally, a relatively large laser power is required to induce thermoelectric effects in a material including MoS2. More specifically, laser power in the range of 100 KW/cm$^2$ (so called high flux regime) is considered necessary to induce observable thermoelectric effects [8-10]. However, we intentionally kept the laser power to two orders of magnitude lower than that of above mentioned high flux regime (< 0.5 KW/cm$^2$) throughout the measurements to prevent a contribution from thermoelectric effects. As a result, the degree of temperature increase by the incident light in our case is estimated to be in the range of few K based on the thermal conductivity of $MoS_2$, and the energy and flux of incident photons. This degree of temperature gradient induced by the incident illumination in SPCM is consistent with literatures [8]. Therefore, the contribution of thermoelectric effects in a discussion of overall photocurrent in our study should be minimal. (2) It is known that a degree of responsivity is much smaller for thermoelectric effects based photocurrent compared to non-thermoelectric based counterpart [9, 10]. The degree of responsivity therefore can be used to determine whether the obtained photocurrent is thermoelectric effects based or not. An estimation of responsivity assuming that our results are thermoelectric effect based photocurrent yields to 1-3 x 10$^{-6}$ A/W at $V_{SD}$ = 0.5 V for both of 2H- and 1T-contacted devices [10]. These values are one and two orders of magnitudes lower than what we observed for 2H- and 1T-contacted devices, respectively; suggesting that our results are not thermoelectric effect based photocurrent.